\documentclass[12pt,a4paper]{article}
\usepackage{amsfonts,latexsym}
\usepackage{amsmath,amssymb}
\usepackage{graphicx,color}

\numberwithin{equation}{section}

\renewcommand{\thefootnote}{\fnsymbol{footnote}}
\newcommand{\nn}{\nonumber}
\begin{document}
%\begin{flushright}
%{\rm CQUeST-2010-xxxx}
%\end{flushright}
\vspace{12mm}

\begin{center}
{{{\Large {\bf Stability of $f(R)$ black holes }}}}\\[10mm]

{Yun Soo Myung$^{a}$\footnote{e-mail address: ysmyung@inje.ac.kr},Taeyoon Moon$^{b}$\footnote{e-mail address: tymoon@sogang.ac.kr},
and
Edwin J. Son$^{b}$\footnote{e-mail address: eddy@sogang.ac.kr}}\\[8mm]

{${}^{a}$ Institute of Basic Sciences and School of Computer Aided Science, Inje University Gimhae 621-749, Korea\\[0pt]}
{${}^{b}$ Center for Quantum Space-time, Sogang University, Seoul, 121-742, Korea\\[0pt] }

\end{center}
\vspace{2mm}

\begin{abstract}
We  investigate the stability of $f(R)$ (Schwarzschild) black hole
obtained from the $f(R)$ gravity. It  is difficult to  carry out the
perturbation analysis around the black hole because the linearized
Einstein equation is fourth order in $f(R)$ gravity. In order to
resolve this difficulty, we transform $f(R)$ gravity into the
scalar-tensor theory by introducing two auxiliary scalars.
 In this case, the linearized curvature scalar becomes a scalaron, showing that all linearized equations are second order,
  which are the  same equations for the massive Brans-Dicke theory.
   It turns out that the  $f(R)$ black hole is stable against the external perturbations if the scalaron does not
have a tachyonic mass.

\end{abstract}
\vspace{5mm}

{\footnotesize ~~~~PACS numbers: }

%{\footnotesize ~~~~Keywords: Classical Theories of Gravity, Spacetime Singularities, Black Holes in String Theory}

\vspace{1.5cm}

\hspace{11.5cm}{Typeset Using \LaTeX}
\newpage
\renewcommand{\thefootnote}{\arabic{footnote}}
\setcounter{footnote}{0}

%%%% Introduction %%%%

\section{Introduction}
Modified gravity theories, $f(R)$ gravities~\cite{NO,sf,NOuh} have
much attentions as one of strong candidates for explaining the
current accelerating universe~\cite{SN}.  $f(R)$ gravities can be
considered as Einstein gravity (massless graviton) with an
additional scalar.   For example, it was shown that the
metric-$f(R)$ gravity is equivalent to the $\omega_{\rm BD}=0$
Brans-Dicke (BD) theory with the potential~\cite{FT}. Although the
equivalence principle test  in the solar system imposes a strong
constraint on $f(R)$ gravities, they may not be automatically ruled
out if the Chameleon mechanism is introduced to resolve it. It was
shown that the  equivalence principle test allows  $f(R)$ gravity
models that are indistinguishable from the ${\rm \Lambda}$CDM model
in the background universe evolution~\cite{PS}. However, this does
not necessarily imply that there is no difference in the dynamics of
perturbations~\cite{CENOZ}.

In order for $f(R)$ gravities to be acceptable, they must obey
certain minimal requirements for theoretical viability~\cite{sf,FT}.
Three important requirements are included: (i) they  possess the
correct cosmological dynamics, (ii) they are free from instabilities
(tachyon) and ghosts~\cite{CZ,CLF,PS}, (iii) they attain the correct Newtonian
and post-Newtonian limits.

On the other hand, the Schwarzschild-de Sitter black hole was
obtained for a positively constant curvature scalar in~\cite{CENOZ}
and other black hole solution  was recently found for a non-constant
curvature scalar~\cite{SZ}.  A black hole solution was obtained from
$f(R)$ gravities by requiring the negative constant curvature scalar
$R=\bar{R}$~\cite{CDM}. If $1+f'(\bar{R})>0$, this black hole is
similar to the Schwarzschild-AdS (SAdS) black hole. In order to
obtain the constant curvature black hole solution from $f(R)$
gravity coupled to the matter,  the trace of its stress-energy
tensor $T_{\mu\nu}$ should be zero.  Hence, two known  matter fields
are the Maxwell~\cite{cdme} and Yang-Mills fields~\cite{MMS}.

 All black hole
solution must pass the stability test. A black hole solution should
be stable against the external perturbations because it stands as
the physically realistic object~\cite{KZ}. There are two ways to
achieve the stability of a black hole: one is the full stability by
considering odd and even perturbations~\cite{Vish} and the other is
the restricted stability by taking into account the spherically
symmetric perturbations for simplicity~\cite{Win}. The latter is not
enough to guarantee the full stability and thus, it must be
supported by the black hole thermodynamics (heat
capacity)~\cite{SYT}. The basic idea is to decouple the linearized
(perturbed) equations and then, manage  to arrive at the second
order Schr\"odinger-type equations for the physical field with the
potential. If all potentials are positive for whole range outside
the event horizon, the black hole under the consideration is stable.
Studies of stability of Kerr black hole are not as straightforward
~\cite{Whi}, because it is axially symmetric black hole and thus,
the decoupling process seems to be complicated.   However, this
method is not suitable for  $f(R)$ black holes because $f(R)$
gravity contains fourth  order derivatives in the linearized
equations~\cite{kerr,BS}. In this case, the requirement (ii) will
play an important role in testing the stability of $f(R)$ black
holes.

In this work, we investigate the stability of $f(R)$ (Schwarzschild)
black hole.  We  transform $f(R)$ gravity into the scalar-tensor
theory to eliminate fourth order derivative terms by introducing two auxiliary scalars. Then, the linearized
curvature scalar  becomes a scalaron, indicating that all linearized
equations are second order. Interestingly, they are exactly the  same
equations for the massive Brans-Dicke theory. Using the stability
analysis of black hole in the massive Brans-Dicke theory, we show
clearly that the  $f(R)$ black hole is stable against the external
perturbations if the scalaron does not have a tachyonic mass.

\section{Perturbation of $f(R)$ black holes}
Let us first consider $f(R)$ gravity without any matter fields whose
action is given by
\begin{eqnarray}
S_{f}=\frac{1}{2\kappa^2}\int d^4 x\sqrt{-g} f(R),\label{Action}
\end{eqnarray}
where $\kappa^2=8\pi G$.  The Einstein equation  takes the form
\begin{eqnarray} \label{equa1}
R_{\mu\nu} f'(R)-\frac{1}{2}g_{\mu\nu}f(R)+
\Big(g_{\mu\nu}\nabla^2-\nabla_{\mu}\nabla_{\nu}\Big)f'(R)=0,
\end{eqnarray}
where ${}^{\prime}$ denotes the differentiation with respect to its
argument.  It is well-known that Eq.(\ref{equa1}) has a solution
with constant curvature scalar $R=\bar{R}$. In this case, Eq.
(\ref{equa1}) can be written as
\begin{eqnarray} \label{equ1}
\bar{R}_{\mu\nu} f'(\bar{R})-\frac{1}{2}g_{\mu\nu}f(\bar{R})=0,
\end{eqnarray}
and thus,  the trace of (\ref{equ1}) becomes
\begin{eqnarray}
\bar{R}f'(\bar{R})-2f(\bar{R})=0.\label{eqR}
\end{eqnarray}
Note that the above equation determines the constant curvature
scalar to be
\begin{eqnarray}
\bar{R}=\frac{2f(\bar{R})}{f'(\bar{R})}\equiv 4\Lambda_f
\label{eqCR}
\end{eqnarray}
with $\Lambda_f$ the cosmological constant due to the $f(R)$
gravity.  Substituting this expression into (\ref{equ1}), one
obtains the Ricci tensor
\begin{equation}
\bar{R}_{\mu\nu}=\frac{f(\bar{R})}{2f'(\bar{R})}\bar{g}_{\mu\nu}=\Lambda_f
\bar{g}_{\mu\nu}.
\end{equation}
The constant curvature black hole solution is given by
\begin{equation} \label{fbh}
ds^2_{\rm
cc}=-\Big(1-\frac{2m}{r}-\frac{\Lambda_f}{3}r^2\Big)dt^2+\frac{dr^2}{1-\frac{2m}{r}-\frac{\Lambda_f}{3}r^2}
+r^2d\Omega^2_2,
\end{equation}
where $\Lambda_f>0, <0,$ =0 denote the Schwarzschild-de Sitter(dS),
Schwarzschild-anti de Sitter (AdS), and Schwarzschild black holes,
respectively. We call these ``$f(R)$ black holes"  because these
were obtained from $f(R)$ gravity. Even though these solutions are
also obtained from the Einstein gravity with cosmological constant,
their perturbation analysis is  different from the Einstein
gravity.   The Schwarzschild-dS black hole has
been extensively studied in $f(R)$ gravities together with the
cosmological implications of $f(R)$ gravities~\cite{CENOZ}.  However, as far as we know, there was a perturbation study on the Schwarzschild black hole in $f(R,G)$ gravities~\cite{FST}.

Now we introduce the perturbation around the constant curvature
black hole to study stability of the black hole
\begin{eqnarray} \label{m-p}
g_{\mu\nu}=\bar{g}_{\mu\nu}+h_{\mu\nu}.
\end{eqnarray}
Hereafter we denote the background quantities with the ``overbar''.
The linearized equation to (\ref{equa1}) is given by (requiring
$f''(\bar{R})\neq0$)
\begin{eqnarray}
f'(\bar{R})\delta
R_{\mu\nu}(h)-\frac{f(\bar{R})}{2}h_{\mu\nu}+f''(\bar{R})
\Bigg[\bar{g}_{\mu\nu}\bar{\nabla}^2-\bar{\nabla}_{\mu}\bar{\nabla}_{\nu}
+\Lambda_f
\bar{g}_{\mu\nu}-\frac{f'(\bar{R})}{2f''(\bar{R})}\bar{g}_{\mu\nu}\Bigg]\delta
R(h)=0,\label{leq}
\end{eqnarray}
where the linearized Ricci tensor and curvature scalar take the
forms
\begin{eqnarray}
\delta
R_{\mu\nu}(h)&=&\frac{1}{2}\Big(\bar{\nabla}^{\rho}\bar{\nabla}_{\mu}h_{\nu\rho}+
\bar{\nabla}^{\rho}\bar{\nabla}_{\nu}h_{\mu\rho}-\bar{\nabla}^2h_{\mu\nu}-\bar{\nabla}_{\mu}
\bar{\nabla}_{\nu}h\Big),\label{lRmunu}\\
\delta R(h)&=&\bar{\nabla}^{\rho}\bar{\nabla}^{\sigma}h_{\rho\sigma}
-\bar{\nabla}^{2}h-\Lambda_f h.\label{lR}
\end{eqnarray}
In order to find the black hole solution, we have to choose a
specific form of $f(R)$ as~\cite{kerr}
\begin{equation}
\label{fform} f(R)=a_1R+a_2R^2+a_3R^3+\cdots.
\end{equation}
We check that $f(0)=0$ at  $R=\bar{R}=0$, which corresponds to
either the Schwarzschild or  Kerr black hole solution. In this work,
for simplicity, we select a spherically symmetric Schwarzschild
black hole solution with $\Lambda_f=0$,
\begin{eqnarray} \label{schw}
ds^2_{\rm Sch}=\bar{g}_{\mu\nu}dx^\mu
dx^\nu=-e^{\nu(r)}dt^2+e^{-\nu(r)}dr^2+r^2(d\theta^2+\sin^2\theta
d\varphi^2)
\end{eqnarray}
with
\begin{equation} \label{schnu}
e^{\nu(r)}=1-\frac{2m}{r}.
\end{equation}
Then, the linearized equation (\ref{leq}) together with
(\ref{lRmunu}) and (\ref{lR}) becomes
\begin{eqnarray}
&&\bar{\nabla}^{\rho}\bar{\nabla}_{\mu}h_{\nu\rho}+
\bar{\nabla}^{\rho}\bar{\nabla}_{\nu}h_{\mu\rho}-\bar{\nabla}^2h_{\mu\nu}
-\bar{\nabla}_{\mu}\bar{\nabla}_{\nu}h
-\bar{g}_{\mu\nu}\Big(\bar{\nabla}^{\alpha}\bar{\nabla}^{\beta}h_{\alpha\beta}
-\bar{\nabla}^2h\Big)\nn\\ &&\hspace*{8em}
+\Big[\frac{2f''(0)}{f'(0)}\Big]\Big(\bar{g}_{\mu\nu}\bar{\nabla}^2-\bar{\nabla}_{\mu}
\bar{\nabla}_{\nu}\Big) \Big(
\bar{\nabla}^{\alpha}\bar{\nabla}^{\beta}h_{\alpha\beta}
-\bar{\nabla}^2h\Big)=0.\label{leq1}
\end{eqnarray}
Taking the trace of $(\ref{leq1})$ with $\bar{g}^{\mu\nu}$,  one
has the fourth order equation for $h_{\mu\nu}$
\begin{eqnarray}
\Big(\bar{\nabla}^2-m^2_f\Big)\Big(
\bar{\nabla}^{\alpha}\bar{\nabla}^{\beta}h_{\alpha\beta}
-\bar{\nabla}^2h\Big)=0\label{leq2}
\end{eqnarray}
with the mass squared $m^2_f$ defined by
\begin{equation}
m^2_f=\frac{f'(0)}{3f''(0)}.
\end{equation}
At this stage, we note that it is not easy to make a further
progress on  the perturbation analysis  because there exist fourth
order derivatives.  We mention  that  for the Einstein gravity with
$f(R)=R$, $f'(R)=1$ and $f''(0)=0$. In this case, one finds  the
equation for linearized curvature scalar: $\delta R(h)=0$, which
means that $\delta R(h)$  is not a physically propagating mode.
Actually, this equation leads to one constraint
\begin{equation} \label{cons}
\bar{\nabla}^{\alpha}\bar{\nabla}^{\beta}h_{\alpha\beta}
=\bar{\nabla}^2h
\end{equation}
which will also be recovered from the transverse gauge. Up to now,
we did not fix any gauge. We would like to comment on the linearized
equation when choosing the Lorentz gauge
\begin{equation}
\bar{\nabla}_{\nu}h^{\mu\nu}=\frac{1}{2}\bar{\nabla}^{\mu}h.
\end{equation}
Under this gauge-fixing, the linearized equation (\ref{leq1})
takes  the  form~\cite{BS}
\begin{eqnarray} \label{bs}
\bar{\nabla}^2\tilde{h}_{\mu\nu}+2\bar{R}_{\mu\rho\nu\sigma}\tilde{h}^{\rho\sigma}
+\frac{f''(0)}{f'(0)}\Big(\bar{g}_{\mu\nu}\bar{\nabla}^2-\bar{\nabla}_{\mu}
\bar{\nabla}_{\nu}\Big)\bar{\nabla}^2\tilde{h}=0
\end{eqnarray}
with the trace-reversed perturbation
$\tilde{h}_{\mu\nu}=h_{\mu\nu}-h\bar{g}_{\mu\nu}/2$~\cite{FH}. This
equation was mainly used to mention that perturbed Kerr black holes
obtained from  $f(R)$ gravity can probe deviations from the Einstein gravity~\cite{BS}.  Even though equation (\ref{bs}) is simpler
than (\ref{leq1}), it is a non-trivial task to decouple odd and even
perturbations around the Kerr black hole, arriving at two fourth
order equations hopefully.  Furthermore, we  do not know how to
solve the fourth order differential equation.  Finally, we may
choose the transverse gauge which   works well for studying the
graviton propagations on the the AdS$_4$ spacetime
background~\cite{GT,Myungf}
\begin{equation}
\bar{\nabla}_{\mu}h^{\mu\nu}=\bar{\nabla}^{\nu}h,
\end{equation}
which leads to (\ref{cons})
 when operating $\bar{\nabla}$ on both sides. Using the
 relation (\ref{cons}), one immediately finds that the effect of $f(R)$ gravity
 [$2f''(0)$-term in (\ref{leq1})] disappears because of $\delta R(h)=0$, leading to the Einstein gravity.
  Hence, the non-covariant gauge-fixing for the black hole perturbation should be
 different from the covariant gauge-fixing for the graviton propagations on
 the AdS, dS and Minkowski spacetimes~\cite{FH}.  It seems that the best way
 to resolve the difficulty confronting with the fourth order
 equation is to translate the fourth order equation into the second
 order equations by introducing auxiliary scalar fields. In other words,
 we must  make a transformation from $f(R)$ gravity to the
 scalar-tensor theory (like Brans-Dicke theory) to analyze the stability of $f(R)$ black hole.

\section{Perturbation of  the scalar-tensor theory}

In this section, we will develop the perturbation analysis around
the $f(R)$ black holes (\ref{fbh}) in the different frame, the
scalar-tensor theory. Introducing two auxiliary fields $\phi$ and
$A$,  one can rewrite the  action (\ref{Action})
as~\cite{Olmo,COT,sf}
\begin{eqnarray}
S_{st}=\frac{1}{2\kappa^2}\int d^4
x\sqrt{-g}\left\{\phi\left(R-A\right)+f(A)\right\}. \label{ActionfA}
\end{eqnarray}
Varying for the fields $\phi$ and $A$ lead to two equations
\begin{eqnarray}
R=A,~~\phi=f'(A).\label{eomA}
\end{eqnarray}
 Note that imposing (\ref{eomA}) on the  action (\ref{ActionfA})
recovers  the original action (\ref{Action}). On the other hand, the
equation of motion for the metric tensor  can be obtained by
\begin{eqnarray} \label{equa}
\phi
R_{\mu\nu}-\frac{f(A)}{2}g_{\mu\nu}
+\Big(g_{\mu\nu}\nabla^2-\nabla_{\mu}\nabla_{\nu}\Big)\phi=0.\label{eomg}
\end{eqnarray}
In deriving the above equation, we used $R=A$ whose reliability was
discussed to explain the solar system  test of $f(R)$
garvity~\cite{sf}. In this work, we use this relation to analyze the
stability of $f(R)$ black hole only.  Considering a constant
curvature scalar $R=\bar{R}=\bar{A}$ together with
$\bar{\phi}=f'(\bar{A})={\rm const}$, Eq.(\ref{eomg}) becomes
\begin{equation}
f'(\bar{A})\bar{R}_{\mu\nu}-\frac{1}{2}\bar{g}_{\mu\nu}f(\bar{A})=0.\label{eomg1}
\end{equation}
Taking the trace of (\ref{eomg1}) leads
to
\begin{eqnarray}
\bar{R}f'(\bar{A})-2f(\bar{A})=0\label{eqR}
\end{eqnarray}
which determines the positive, negative and zero curvature scalar by choosing a form of $f(A)$
\begin{eqnarray}
\bar{R}=\frac{2f(\bar{A})}{f'(\bar{A})}\equiv 4\Lambda_A.
\label{eqCR}
\end{eqnarray}
Substituting this expression into (\ref{eomg1}), one finds the
Ricci tensor which determines the maximally symmetric Einstein spaces including Minkowski space
\begin{equation}
\bar{R}_{\mu\nu}=\frac{1}{2}\frac{f(\bar{A})}{f'(\bar{A})}\bar{g}_{\mu\nu}=\Lambda_A
\bar{g}_{\mu\nu}.
\end{equation}
Now we are in a position to  study the perturbation around the
constant curvature
 black hole (\ref{fbh}).  In addition to (\ref{m-p}), from (\ref{eomA}), we have
\begin{equation}
\bar{R}+\delta R(h)=\bar{A}+\delta A,~~\bar{\phi}+\delta \phi=f'(\bar{A})+f''(\bar{A})\delta A,
\end{equation}
which leads to
\begin{equation}
\delta R(h)\to \delta A,~~\delta \phi \to f''(\bar{A})\delta A.
\end{equation}
Thus,  instead of $\delta R(h)$ and $\delta \phi$, we use  $\delta
A$ as a perturbed field  in addition to $h_{\mu\nu}$. We expect that
the same results  can be derived when using $h_{\mu\nu}$ and $\delta
\phi$ in the Brans-Dicke theory because $\delta \phi \simeq \delta
A$.

The linearized equation to (\ref{eomg}) takes the form
\begin{eqnarray}
\delta R_{\mu\nu}(h)&-&\Lambda_A
h_{\mu\nu}+\bar{g}_{\mu\nu}\Bigg[\frac{f''(\bar{A})f(\bar{A})-f'^{2}(\bar{A})}{2f'^{2}(\bar{A})}\bigg]\delta
A \nonumber \\
\label{pertg} &+&\Big[\frac{f''(\bar{A})}{f'(\bar{A})}\Big]
\Big(\bar{g}_{\mu\nu}\bar{\nabla}^2-\bar{\nabla}_{\mu}\bar{\nabla}_{\nu}\Big)\delta
A=0,
\end{eqnarray}
where the linearized Ricci tensor  $\delta R_{\mu\nu}(h)$ is given
by (\ref{lRmunu}).  It is important to note that taking the trace of
$(\ref{pertg})$ with $\bar{g}^{\mu\nu}$ leads to  the linearized
second order ``scalaron" equation, instead of the linearized fourth
order curvature scalar equation (\ref{leq2}),  as
\begin{eqnarray}
\Big(\bar{\nabla}^2-m_A^2\Big)\delta A=0,\label{gtr}
\end{eqnarray}
where the scalaron mass squared $m_A^2$ is given by
\begin{eqnarray}
m_A^2=\frac{f'^{2}(\bar{A})-2f(\bar{A})f''(\bar{A})}{3f'(\bar{A})f''(\bar{A})}
=\frac{f'(\bar{A})}{3f''(\bar{A})}-\frac{4}{3}\Lambda_A,\label{mass}
\end{eqnarray}
which was already known as (97) of~\cite{sf} in  dS spacetimes. This
is the main reason  why we have introduced the action
(\ref{ActionfA}), instead of (\ref{Action}). Plugging Eq.(\ref{gtr})
into Eq.(\ref{pertg}) and rearranging the terms,  we arrive at the
linearized second order equation
\begin{eqnarray}
\delta R_{\mu\nu}(h)-\Lambda_A
h_{\mu\nu}=\Big[\frac{f''(\bar{A})}{f'(\bar{A})}\Big]\bar{\nabla}_{\mu}\bar{\nabla}_{\nu}\delta
A+\bar{g}_{\mu\nu}\Bigg[\frac{f'^{2}(\bar{A})+f(\bar{A})f''(\bar{A})}{6f'^{2}(\bar{A})}\Bigg]
\delta A. \label{eqRmunu}
\end{eqnarray}
In this work,  we confine ourselves to the asymptotically flat
spacetimes  with $\Lambda_A=0$ which accommodates the Schwarzschild
black hole. Also, we do not choose an explicit form of $f(A)$ since
such a restriction seems  unnecessary to study the stability of
$f(R)$ black holes.   In this case, we have to choose $\bar{A}=0$
and thus,
\begin{equation}
f(0)=0,~~ f'(0)\neq 0,~~f''(0)\neq 0 \label{mink}.
\end{equation}
Taking into account (\ref{mink}),
Eq.(\ref{eqRmunu}) reduces to
\begin{eqnarray}
\delta
R_{\mu\nu}(h)=\Big[\frac{1}{3m^2_A}\Big]\bar{\nabla}_{\mu}\bar{\nabla}_{\nu}\delta
A+\frac{1}{6}\bar{g}_{\mu\nu}\delta A\label{eqRmunu1},
\end{eqnarray}
where the scalaron mass squared in the asymptotically flat
spacetimes is given by
\begin{equation}
m^2_A=\frac{f'(0)}{3f''(0)}.
\end{equation}
Since the mass dimension of  the linearized scalaron is two
($[\delta A]=2$), it would be better to write the canonical
linearized equations by introducing a dimensionless scalaron $\delta
\tilde{A}$ defined by
\begin{equation}
\delta \tilde{A}=\frac{\delta A}{3m^2_A}. \end{equation} Finally, we
arrive at two linearized equations
\begin{eqnarray}
&&\Big(\bar{\nabla}^2-m_A^2\Big)\delta \tilde{A}=0,\label{gtrf} \\
&& \delta R_{\mu\nu}(h)-\bar{\nabla}_{\mu}\bar{\nabla}_{\nu}\delta
\tilde{A}-\Big[\frac{m^2_A}{2}\Big]\bar{g}_{\mu\nu}\delta
\tilde{A}=0, \label{eqRmunu1f} \end{eqnarray} which are our main
result for carrying out the stability analysis of $f(R)$ black hole.
Importantly, we observe that when replacing
 \begin{equation}
 \delta \tilde{A} \to \varphi,~~ m^2_A \to -c,
 \end{equation}
 Eqs.(\ref{gtrf}) and (\ref{eqRmunu1f}) are exactly the same equations of  the
massive Brans-Dicke theory for the  stability analysis of the
Schwarzschild black hole~\cite{pert}. We mention that the stability
analysis for the Schwarzschild black hole in the Brans-Dicke theory
without potential ($c=0$) has been established in~\cite{kkmcp}.
Recently, the scalar field perturbations of Schwarzschild black hole
was carried out in tensor-vector-scalar  theory~\cite{LD}.  Hence,
we will use the result for the stability analysis for the
Schwarzschild black hole in the massive Brans-Dicke theory
($c\not=0$).

Finally, we would like to mention that even though the Brans-Dicke
 scalar  was introduced instead of scalaron $\delta A$, there is no
 change in the linearized equations. (See Appendix).
 For this purpose, we rewrite (\ref{eqRmunu}) as
\begin{equation}
\delta R_{\mu\nu}(h)-\Lambda_A
h_{\mu\nu}-\bar{\nabla}_{\mu}\bar{\nabla}_{\nu}\delta
\tilde{A}-\Big[\frac{m^2_A}{2}+\Lambda_A\Big]\bar{g}_{\mu\nu}\delta
\tilde{A}=0. \label{23eq}
\end{equation}
 The only
 difference seems to be the mass squared $m^2_\phi$, in compared to
 $m^2_A$ in (\ref{mass}). However, they become the same mass squared
 when making replacements (\ref{replace}).

\section{Stability analysis of $f(R)$ black hole \\
in the scalar-tensor theory}

The metric perturbations $h_{\mu\nu}$ are classified depending on
the transformation properties under parity, namely odd (axial) and
even (polar). Using the Regge-Wheeler~\cite{regge}, and Zerilli
gauge~\cite{Zeri} , one obtains two distinct perturbations : odd and
even perturbations. For odd parity, one has with two off-diagonal
components $h_0$ and $h_1$
\begin{eqnarray}
h^o_{\mu\nu}=\left(
\begin{array}{cccc}
0 & 0 & 0 & h_0(r) \cr 0 & 0 & 0 & h_1(r) \cr 0 & 0 & 0 & 0 \cr
h_0(r) & h_1(r) & 0 & 0
\end{array}
\right) e^{-ikt}\sin\theta\frac{dp_{l}}{d\theta} \,, \label{oddp}
\end{eqnarray}
while for even parity, the metric tensor takes the form with four components $H_0,~H_1,~H_2,$ and $K$ as
\begin{eqnarray}
h^e_{\mu\nu}=\left(
\begin{array}{cccc}
H_0(r) e^{\nu(r)} & H_1(r) & 0 & 0 \cr H_1(r) & H_2(r) e^{-\nu(r)} &
0 & 0 \cr 0 & 0 & r^2 K(r) & 0 \cr 0 & 0 & 0 & r^2\sin^2\theta K(r)
\end{array}
\right) e^{-ikt}p_{l} \,, \label{evenp}
\end{eqnarray}
where $p_l$ is Legendre polynomial with angular momentum $l$ and
$e^{\nu(r)}$ was given by (\ref{schnu}). We note that (\ref{oddp})
and (\ref{evenp}) correspond to the non-covariant
gauge-fixing~\cite{CB}.

 In
order to explain the stability analysis of $f(R)$ black hole
briefly, we mainly use the result for the stability analysis for the
massive Brans-Dicke theory~\cite{pert}.  For the odd-parity
perturbation, its linearized equation takes a simple form as
\begin{equation}
\delta R_{\mu\nu}(h)=0
\end{equation}
which leads to the Regge-Wheeler equation by introducing the tortoise coordinate $r^*=r+2m\ln[r/2m-1]$
\begin{equation}
\frac{d^2Q}{dr^{*2}}+\Big[k^2-V_{RW}\Big]Q=0,
\end{equation}
where the Regge-Wheeler potential is given by~\cite{regge}
\begin{equation}
V_{RW}(r)=\Big(1-\frac{2m}{r}\Big)\Big[\frac{l(l+1)}{r^2}-\frac{6m}{r^3}\Big].
\end{equation}
This potential is always positive for whole range of $-\infty
<r^*<\infty$  and  a barrier-type localized around $r^*=0$, (see
Fig. 3 of~\cite{kkmcp}) which implies that the odd-perturbation is
stable~\cite{Vish}. For the even-perturbation, we have to use the
linearized equation (\ref{eqRmunu1f})
 because the scalaron $\delta \tilde{A}$ contributes to making an even mode $\hat{M}$ together with $H_0,~H_1,~H_2,$
 and $K$ definitely. After a long algebraic manipulation, we arrive at the Zerilli's equation
\begin{equation}
\frac{d^2\hat{M}}{dr^{*2}}+\Big[k^2-V_{Z}\Big]\hat{M}=0,
\end{equation}
where the Zerilli potential is given by~\cite{Zeri}
\begin{equation}
V_{Z}(r)=\Big(1-\frac{2m}{r}\Big)\Bigg[\frac{2\lambda^2(\lambda+1)r^3+6\lambda^2mr^2+18\lambda
m^2 r+18m^3} {r^3(\lambda r+3m)^2}\Bigg]
\end{equation}
with
\begin{equation}
\lambda=\frac{1}{2}(l-1)(l+2).
\end{equation}
The Zerilli potential $V_{Z}$ is always positive for whole range of
$-\infty <r^*<\infty$  and  a barrier-type localized around $r^*=0$
(see Fig. 3 of~\cite{kkmcp}), which implies that the
even-perturbation is stable, even though the scalaron $\delta
\tilde{A}$ is coupled to the even-parity perturbations.

 Finally, when considering
 \begin{equation}
 \delta \tilde{A} \propto \Sigma \frac{\psi(r)}{r} Y_{lm}(\theta,\varphi) e^{-ikt},
 \end{equation}
 the linearized  scalaron equation (\ref{gtrf}) leads to the Schr\"odinger-type equation
 \begin{equation}
\frac{d^2\psi}{dr^{*2}}+\Big[k^2-V_{A}\Big]\psi=0,
\end{equation}
where the scalaron  potential is given by
\begin{equation}
V_{A}(r)=\Big(1-\frac{2m}{r}\Big)\Big[\frac{l(l+1)}{r^2}+\frac{2m}{r^3}+m^2_A\Big],
\end{equation}
where the second term is the usual scalar term with spin zero [in
general, $-2m(s^2-1)$ for $s$ spin-weight of the perturbing field],
while the last term shows clearly the feature of a massive scalaron
arisen from  $f(R)$ gravity. The potential $V_A(r)$ is always
positive  exterior the event horizon if the mass squared $m^2_A$ is
positive (non-tachyonic mass).  For the shapes of scalaron
potential, see Fig. 2 and Fig. 3 of~\cite{pert} by replacing $m^2_A$
by $-c$. However, if $m^2_A$  is negative (tachyonic mass), the
potential becomes negative for large $r$ and approaches $-|m^2_A|$
at infinity, indicating that the $f(R)$ black hole is unstable
against the scalaron-perturbation.

\section{Discussions}

We have investigated the stability of $f(R)$ (Schwarzschild) black
hole obtained from the $f(R)$ gravity. Actually, it seems to be  a
formidable task to carry out the perturbation analysis around the
black hole  because the linearized Einstein equation is fourth order
in $f(R)$ gravity.

We have proposed that  the best way
 to resolve the difficulty confronting with the fourth order
differential  equation is to translate the fourth order equation
into the second
 order equation by introducing auxiliary scalar fields.
 In this case, the linearized curvature scalar  $\delta R(h)$ becomes a massive scalaron, showing that all linearized equations are second
 order. We observed that the canonical linearized equations
 become  the  same equations for the massive Brans-Dicke theory when replacing the scalaron (its mass $m^2_A$) by the Brans-Dicke scalar (its mass $-c$).
 Then, it is straightforward to make a decision on the stability of
 $f(R)$ black hole.

 The stability on the metric perturbations remains
 unchanged,  confirming that the odd (even) perturbations lead to
 the Schr\"odinger-type equation with the Regge-Wheeler (Zerilli)
 potential. This corresponds to the Einstein gravity, even though
 the even mode contains the scalaron in addition to $H_0,~H_1,~H_2,$ and
 $K$.
The difference comes from the linearized scalaron equation
 because the scalaron is a massive scalar which is physically  propagating on the black
 hole  background.
It turns out that the $f(R)$  is stable against the external
perturbations if the scalaron does not have a tachyonic mass
($f''(0)>0$). This is consistent with other perturbation analysis:
the Dolgov-Kawasaki instability with $f''(R)<0$ in cosmological
perturbations~\cite{DK}, graviton and scalar propagations in the
Minkowski~\cite{Olmo}, dS~\cite{fn,olmod} and AdS~\cite{Myungf}
spacetimes.

 We would like to compare our stability analysis with ref.\cite{CDM},
 where the very restricted stability was performed  by taking into account the spherically
symmetric static  perturbations for simplicity. Therefore, their
results did not show the full stability analysis, in contrast with
our results.

In this work, even though the perturbation formalism is suitable for
all constant curvature black holes (\ref{fbh}), including the Kerr
black hole, we have analyzed the Schwarzschild black hole.  We
conjecture that the stability of black holes from Einstein gravity
theory may hold for the $f(R)$ black holes if one uses the
scalar-tensor approach developed in Section 3.

 \vspace{1cm}
{\bf Acknowledgments}

This work was supported by the 2010 Inje University Research Grant.

\section*{Appendix: Brans-Dicke approach }

We may rewrite (\ref{ActionfA}) as the Brans-Dicke  theory
\begin{eqnarray}
S_{\phi}=\frac{1}{2\kappa^2}\int d^4 x\sqrt{-g}\left\{\phi
R-V(\phi)\right\} \label{Actionfphi}
\end{eqnarray}
with the potential
\begin{equation}
V(\phi)=\phi A(\phi)-f(A(\phi)).
\end{equation}
Varying for the fields $g_{\mu\nu},~\phi$ lead to the following
equations:
\begin{eqnarray}
&&\phi\left(R_{\mu\nu}-\frac{1}{2}g_{\mu\nu}R\right)+\frac{1}{2}g_{\mu\nu}V(\phi)
+\Big(g_{\mu\nu}\nabla^2-\nabla_{\mu}\nabla_{\nu}\Big)\phi,\label{eomg}\\
&&R=V'(\phi) \label{eomphi}
\end{eqnarray}
where ${}^{\prime}$ denotes differentiation with respect to $\phi$.
 Taking the trace  of (\ref{eomg}) in order to replace $R$ in
 (\ref{eomphi}) leads to the scalar equation
 \begin{equation}
 3\nabla^2\phi+2V(\phi)-\phi V'(\phi)=0.
 \end{equation}
For the constant curvature scalar case only, we have
\begin{equation}\phi=\bar{\phi},~~
\bar{R}=V'(\bar{\phi})=\frac{2V(\bar{\phi})}{\bar{\phi}}=4\Lambda_\phi,~~\bar{R}_{\mu\nu}=\Lambda_\phi
g_{\mu\nu}\label{consphi}
\end{equation}
where
\begin{equation}
\Lambda_\phi=-\frac{3}{\ell^2}
\end{equation}
 with $\ell$ the AdS$_4$ curvature radius.
  From (\ref{eomphi}) and
(\ref{consphi}),  the potential may take the form
\begin{equation}
V_\phi=c_0+c_1(\phi-\bar{\phi})+c_2(\phi-\bar{\phi})^2+\cdots,
\end{equation}
where
\begin{equation}
c_0=\frac{c_1}{2}\bar{\phi}<0,~~c_1<0,~~c_2>0,~~\bar{\phi}>0.
\end{equation}
 The linearized equations
around the constant curvature scalar background (\ref{consphi})  can
be obtained as
\begin{eqnarray}
&&\Big(\bar{\nabla}^2-m_{\phi}^2\Big)\delta \tilde\phi=0,\label{gt11} \\
&& \delta R_{\mu\nu}(h)-\Lambda_{\phi}
h_{\mu\nu}-\bar{\nabla}_{\mu}\bar{\nabla}_{\nu}\delta
\tilde\phi-\Big[\frac{m^2_{\phi}}{2}+\Lambda_{\phi}\Big]\bar{g}_{\mu\nu}\delta
\tilde\phi=0, \label{eq12phi} \end{eqnarray} where
$\delta\tilde\phi=\delta\phi/\bar\phi$ and $m_{\phi}^2$ is given by
\begin{equation}
m_{\phi}^2=\frac{1}{3}\left(\bar\phi
V''(\bar{\phi})-V'(\bar{\phi})\right),\label{phimass}
\end{equation}
where
\begin{equation}
\bar{\phi} V''(\bar{\phi})=2c_2\bar{\phi}>0.
\end{equation}
We emphasize that the mass squared (\ref{phimass}) is exactly the
same as one derived in the literature \cite{Olmo,Olmof} with zero
Brans-Dicke parameter ($\omega=0$). Also, it is easily shown that
when making the replacements  \begin{equation} \label{replace}
\delta \tilde{\phi} \to \delta \tilde{A} ,~\bar{\phi} \to
f'(\bar{A}),~V(\bar{\phi}) \to f(\bar{A}),~V'(\bar{\phi}) \to
\bar{A},~V''(\bar{\phi}) \to \frac{1}{f''(\bar{A})},
\end{equation}
(\ref{gtrf}) and (\ref{23eq}) lead to (\ref{gt11}) and
(\ref{eq12phi}), respectively. Hence, it is enough to solve
(\ref{gtrf}) and (\ref{eqRmunu1f}) for the stability analysis of
$f(R)$-Schwarzschild black holes with $\Lambda_A=0$.

\end{document}